\documentclass[%
 reprint,
 amsmath,amssymb,
 aps,
floatfix,
]{revtex4-1}

\usepackage{graphicx}
\usepackage{dcolumn}
\usepackage{bm}

\newcommand{\pom}{I\!\!P}
\def\beq{\begin{equation}}
\def\eeq{\end{equation}}
\def\bea{\begin{eqnarray}}
\def\eea{\end{eqnarray}}

\parskip=\itemsep

\def\pom{{I\!\!P}}
\begin{document}
\date{\today}
\title
{\bf \Large
Unitarity Saturation In P-P Scattering}
\author
{\bf \large
Uri Maor}
\affiliation
{School of Physics and Astronomy\\
Raymond and Beverly Sackler Faculty of Exact Science\\
Tel Aviv University, Tel Aviv, 69978, Israel.}
\begin{abstract}
The properties of soft p-p scattering amplitudes at the TeV-scale
are studied so as to identify the impact of s and t channel
unitarity screenings on their behavior at exceedingly high energies
and determine the rate at which they approach the bounds
implied by unitarity saturation.
I shall examine the relevant high energy soft cross section
features, as well as, the corresponding behavior of the coupled
phenomenological models aiming to reproduce this data.\\
My conclusion is that p-p black body saturation is not attained up to 
100 TeV. More over, I do not expect that saturation will be
attained at energies that can be investigated experimentally.
\end{abstract}
\maketitle
\section{INTRODUCTION}
Following are 3 paradoxes, dating back to the ISR epoch, which are
resolved by the introduction of unitarity screenings.
\begin{itemize}
\item
Whereas non screened $\sigma_{tot}$ grows like $s^{\Delta}$,
$\sigma_{el}$ grows faster, like $s^{2\Delta}$
(up to logarithmic corrections).
With no screening, $\sigma_{el}$
will, eventually, be larger than $\sigma_{tot}$.
\item
Elastic and diffractive scatterings are seemingly
similar. However, the
energy dependence of the diffractive cross sections is
significantly more moderate than that of $\sigma_{el}$.
\item
The elastic amplitude is central in impact parameter b-space, peaking
at b=0. The diffractive amplitudes are peripheral peaking
at large b, which gets larger with energy.
\end{itemize}
In the following I wish to explore the features of 
elastic scattering and inelastic diffractive scattering
and their impact on our investigation of unitarity screenings. 
As we shall see, 
models confined only to elastic scattering are single dimension.
Incorporating
diffraction in our formalism implies a two dimension presentation
of the unitarity equation. Recall, though, that enforcing unitarity 
is model dependent.\\
Added to our data analysis, is the output of two updated versions
of the Pomeron ($\pom$) model.
Regardless of their differences, GLM and KMR models
provide compatible procedures to calculate both s and t channel screenings
of elastic and diffractive scattering. The two models have
a single partonic Pomeron. Its hardness
depends on the $\pom$ screenings(GLM),
or the transverse momenta of its partons(KMR).\\
Current $\pom$ models have a relatively
large $\Delta_{\pom}$ and exceedingly small (non zero)   
$\alpha_{\pom}^{\prime}$, which seemingly disagree with the conventional
features of the Regge Pomeron, in which the s dependence of a $\pom$
exchange amplitude is determined by $\Delta_{\pom}$ and
the shrinkage of its forward t slope by $\alpha_{\pom}^{\prime}$.
In the $\pom$ models,
the traditional Regge features are restored by
s and t unitarity screenings.\\
Both GLM and KMR utilize the approximation
$\alpha_{\pom}^{\prime}=0$.   
This assumption is critical for the input of a single Pomeron,
and the summation of higher order $\pom$ diagrams. It
implies an upper validity bound 
of these models at 60-100 TeV.\\
Since I wish to assess
unitarity saturation also above 100 TeV,
I have included in the
analysis also the Block-Halzen calculations of
the total and inelastic cross sections in a single channel model
based on a logarithmic parametrization.
This model can be applied at arbitrary high energies.
Recall that, single channel models are deficient 
since they neglect the diffractive channels.\\
This talk aims to assess the approach of p-p scattering 
amplitudes toward s and t
unitarity saturation.
The analysis I shall present is based on:
\begin{itemize}
\item
General principles manifested by Froissart-Martin
asymptotic bound of p-p
total cross sections, introduced 50 years ago.
\item
TeV-scale p-p data analysis based on the output of
the TEVATRON, LHC, and AUGER (in which
p-p features are calculated from
p-Air Cosmic Rays data).
\item
As we shall see,
the TEVATRON(1.8)-LHC(7)-AUGER(57) data
indicate that soft scattering amplitudes
populate a small, slow growing, fraction of the
available phase space confined by unitarity bounds.
\item
Phenomenological unitarity models substantiate the conclusions
obtained from the available data analysis. 
Model predictions suggest
that saturation is attained (if at all) at much higher energies 
well above experimental reach.
\end{itemize}
\section{S CHANNEL UNITARITY}
The simplest s-channel unitarity bound on $a_{el}(s,b)$
is obtained from a diagonal re-scattering matrix,
where repeated elastic re-scatterings secure s-channel unitarity:
\begin{equation}
2Im a_{el}(s,b)={\mid}a_{el}(s,b){\mid}^2+G^{in}(s,b).
\end{equation}
Its general solution is
\begin{eqnarray}
a_{el}(s,b)\,&=&\,i\left(1-e^{-\Omega(s,b)/2}\right), \nonumber \\
G^{in}(s,b)\,&=&\,1\,- e^{-\Omega(s,b)}.
\end{eqnarray}
$\Omega$ is arbitrary.
The output s-unitarity bound is $\mid a_{el}(s,b)\mid\leq 2$,
leading to very large total  
and elastic LHC cross sections, which are
not supported by the recent Totem data.\\
\begin{figure}
\includegraphics[width=9cm,height=8cm]{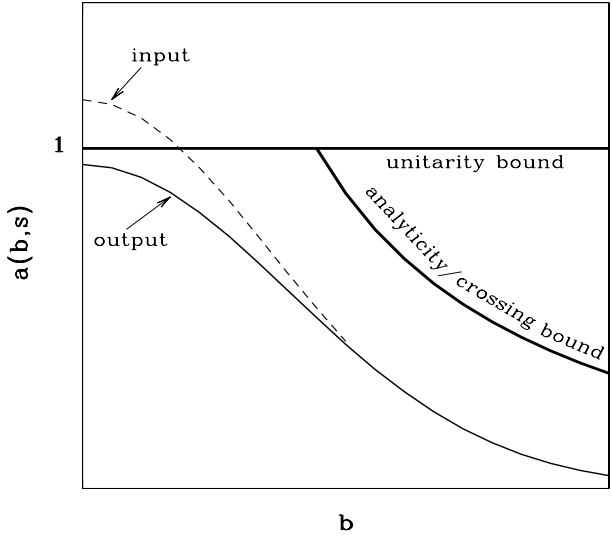}   
\caption{The effect of eikonal screening restoring s-unitarity.  
The schematic bound implied by analyticity/crossing is also shown.}
\end{figure}
In a Glauber type eikonal approximation, the
input opacity $\Omega(s,b)$ is real.
It equals to the imaginary part of the input Born term,
a $\pom$ exchange in our context.
The output $a_{el}(s,b)$ is imaginary.\\
The consequent bound is $\mid a_{el}(s,b)\mid\leq 1,$
which is the black disc bound.\\
Analyticity and crossing symmetry
are restored by the dispersion relation substitution
\begin{equation}
s^{\alpha_{\pom}}\,
\rightarrow\,s^{\alpha_{\pom}}e^{-\frac{1}{2}i\pi \alpha_{\pom}}.
\end{equation}
In a single channel eikonal model, the screened cross sections are:
\begin{eqnarray}
\sigma_{tot}&=&2\displaystyle\int d^2 b \left(1-e^{-\Omega(s,b)/2}\right), \\
\sigma_{el}&=&\displaystyle\int d^2 b
\left(1-e^{-\Omega(s,b)/2}\right)^2, \\
\sigma_{inel}&=&\displaystyle\int d^2 b \left(1-e^{-\Omega(s,b)}\right).
\end{eqnarray}
An illustration of the effects implied by unitarity screenings
are shown in Fig.1. It shows the s-channel black bound of unity,
and the bound implied by analyticity/crossing symmetry on the 
expanding b-amplitude.\\
Imposing these limits leads to the Froissart-Martin bound:
\begin{eqnarray}
\sigma_{tot}\,&\leq&\,C ln^2(s/s_0), \nonumber \\
s_0&=&1GeV^2,\nonumber \\
C\,&\propto&\,1/2m^2_{\pi}\simeq 30mb.
\end{eqnarray}
C is far too large to be relevant in the analysis of
TeV-scale data.\\  
Coupled to Froissart-Martin is MacDowell-Martin bound: $\,\,$
$\frac{\sigma_{tot}}{B_{el}}\,\leq\,18\,\pi\,
\frac{\sigma_{el}}{\sigma_{tot}}.$\\
Note that the Froissart limit controls the asymptotic behavior of the
unitarity cross section bound, 
NOT the behavior of the elastic scattering cross section as such, which
can have an arbitrary functional behavior as long as it is bellow
saturation.\\
There have been recent suggestions by Azimov,
Fagundes et al., and Achilli et al.,
to revise the normalization and/or the functional behavior of 
the bound. As it stands, these attempts are not relevant to
our analysis.\\
In t-space, $\sigma_{tot}$ is proportional to a single point, 
$d\sigma_{el}/dt(t=0)$ (optical theorem).\\ 
As we saw, $\sigma_{tot}$ in b-space is obtained from a 
$b^2$ integration over $2(1-e^{-\frac{1}{2}\Omega(s,b)}).$\\ 
Saturation in b-space is, thus, a differential feature, 
attained initially at b=0  
and then expands very slowly with energy.
Consequently, a black core is a product of partial saturation, 
different from a complete saturation in which $a_{el}(s,b)$ 
is saturated at all b.\\
In a single channel model, 
$\sigma_{el} \leq \frac{1}{2}\sigma_{tot}$ and 
$\sigma_{inel} \geq \frac{1}{2} \sigma_{tot}.$\\ 
At saturation, regardless of the energy at which it is attained, 
\begin{equation}
\sigma_{el} = \sigma_{inel} = \frac{1}{2} \sigma_{tot}. 
\end{equation}
Introducing diffraction, 
will significantly change the features of unitarity 
screenings. However, the saturation signatures remain valid. 
\section{TEV-SCALE DATA}
Following is p-p TeV-scale data relevant to 
the assessment of saturation:\\
CDF(1.8 TeV): 
\begin{eqnarray}
\sigma_{tot}&=&80.03 \pm 2.24 mb, \nonumber \\
\sigma_{el}&=&19.70 \pm 0.85 mb, \nonumber \\
B_{el}&=&16.98 \pm 0.25 GeV^{-2}.\nonumber  
\end{eqnarray}
TOTEM(7 TeV): 
\begin{eqnarray}
\sigma_{tot}&=&98.3 \pm 0.2(stat) \pm 2.8(sys) mb,\nonumber \\
\sigma_{el}&=&24.8 \pm 0.2(stat) \pm 1.2(sys) mb,\nonumber \\
B_{el}&=&20.1 \pm 0.2(stat) \pm 0.3(sys) GeV^{-2}. \nonumber 
\end{eqnarray}
AUGER(57 TeV):
\begin{eqnarray}
\sigma_{tot}&=&133 \pm 13(stat) {\pm}^{17}_{20}(sys) \pm 16(Glauber) mb,
\nonumber \\
\sigma_{inel}&=&92 \pm 7(stat) \pm^{9}_{11}(sys) \pm 16(Glauber) mb.
\nonumber
\end{eqnarray}
Consequently:
\begin{eqnarray}
\sigma_{inel}/\sigma_{tot}(CDF)\,&=&\,0.75, \nonumber \\
\sigma_{inel}/\sigma_{tot}(TOTEM)\,&=&\,0.75, \nonumber \\
\sigma_{inel}/\sigma_{tot}(AUGER)\,&=&\,0.69.\nonumber \\
\sigma_{tot}/B_{el}(TOTEM)&=&12.6<14.1.\nonumber    
\end{eqnarray}
The ratios above imply that saturation of the
elastic p-p amplitude has NOT been attained up to 57 TeV.
Note that the margin of AUGER errors is large.
Consequently, saturation studies in the TeV-scale need the support of 
phenomenological models!
\section{POMERON MODEL}
Translating the concepts presented into a viable phenomenology
requires a specification of $\Omega(s,b)$, for which
Regge theory is a powerful tool. 
Pomeron ($\pom$) exchange is the leading term 
in the Regge hierarchy.\\
The growing total and elastic 
cross sections in the ISR-Tevatron range are well reproduced by
the non screened single channel DL $\pom$ model in which:
\begin{eqnarray} 
\alpha_{\pom}(t)&=&1+\Delta_{\pom}+\alpha^{\prime}_{\pom}t, \nonumber \\
\Delta_{\pom}&=&0.08, \nonumber \\
\alpha^{\prime}_{\pom}&=&0.25GeV^{-2}.
\end{eqnarray}
$\Delta_{\pom}$ determines the energy dependence,  
and $\alpha^{\prime}_{\pom}$ the forward slopes.\\
Regardless of DL remarkable success at lower energies, 
they under estimate the LHC cross sections. 
This is traced to  
DL neglect of diffraction and unitarity screenings
initiated by s and t dynamics. 
Updated Pomeron models analyze elastic and diffractive channels 
utilizing s and t unitarity screenings.
\subsection{Good-Walker Decomposition}
Consider a system of two orthonormal states,
a hadron $\Psi_h$  and a diffractive state $\Psi_D$. 
$\Psi_D$ replaces the continuous 
diffractive Fock states. 
Good-Walker (GW) noted that  
$\Psi_h$ and $\Psi_D$ do not diagonalize 
the 2x2 interaction matrix ${\bf T}$.\\
Let $\Psi_1$, $\Psi_2$
be eigen states of $\bf T.$  
\begin{eqnarray}
&& \Psi_h = \alpha\,\Psi_1 + \beta\,\Psi_2,\nonumber \\
&& \Psi_D = -\beta\,\Psi_1 + \alpha\,\Psi_2, \nonumber \\
&& \alpha^2 + \beta^2 = 1,
\end{eqnarray}  
initiating 4 $A_{i,k}$ elastic GW amplitudes 
$(\psi_i+\psi_k \rightarrow \psi_i+\psi_k).$ i,k=1,2.\\ 
For initial $p({\bar p})-p$ we have $A_{1,2}=A_{2,1}$. 
I shall follow the GLM 
definition, in which the mass 
distribution associated with $\Psi_D$ is not defined.\\ 
The elastic, SD and DD 
amplitudes in a 2 channel GW model are:
\begin{equation} 
a_{el}(s,b)=i\{\alpha^4A_{1,1}+2\alpha^2\beta^2A_{1,2}+\beta^4A_{2,2}\}, 
\end{equation}
\begin{equation}
a_{sd}(s,b)=i\alpha\beta\{-\alpha^2A_{1,1}+ 
(\alpha^2-\beta^2)A_{1,2}+\beta^2A_{2,2}\},
\end{equation}
\begin{equation}
a_{dd}(s,b)=i\alpha^2\beta^2\{A_{1,1}-2A_{1,2}+A_{2,2}\}. 
\end{equation}
\begin{equation}
A_{i,k}(s,b)=\left(1-e^{\frac{1}{2}\Omega_{i,k}(s,b)}\right)\leq 1.
\end{equation}
GW mechanism changes the structure of s-unitarity below saturation.
\begin{itemize}
\item
In the GW sector we obtain the Pumplin bound:
$\sigma_{el}+\sigma_{diff}^{GW} \leq \frac{1}{2}\sigma_{tot}$.\\
$\sigma_{diff}^{GW}$ is the sum of the GW soft diffractive 
cross sections. 
\item
Below saturation,
$\sigma_{el} \leq \frac{1}{2}\sigma_{tot}-\sigma_{diff}^{GW}$
and $\sigma_{inel} \geq \frac{1}{2} \sigma_{tot}+\sigma_{diff}^{GW}.$
\item
$a_{el}(s,b)=1,$ when and only when,
$A_{1,1}(s,b)=A_{1,2}(s,b)=A_{2,2}(s,b)=1$.
\item
When $a_{el}(s,b)=1,$ all diffractive amplitudes at (s,b) vanish.
\item
As we shall see, there is a distinction between  
GW and non GW diffraction. 
Regardless, 
GW saturation signatures are 
valid also in the non GW sector.
\item 
As we saw, the saturation signature,
$\sigma_{el}=\sigma_{inel}=\frac{1}{2}\sigma_{tot},$ 
in a multi channel calculation is coupled to 
$\sigma_{diff}=0.$ Consequently, 
prior to saturation the diffractive cross sections 
stop growing and start to decrease with energy.
This is a clear signature preceding saturation.
\end{itemize}
\section{CROSSED CHANNEL UNITARITY}
{\begin{figure*}
\includegraphics[width=12cm,height=5cm]{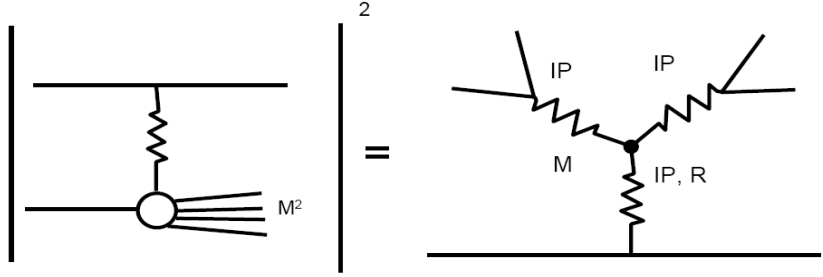}
\caption{Mueller's triple Regge diffractive diagram}
\end{figure*}}
Mueller(1971) applied 3 body unitarity to equate the cross section of
$a\,+\,b \rightarrow M_{sd}^2\,+\,b$ to the triple 
Regge diagram 
$a\,+b\,+\bar{b} \rightarrow a\,+\,b\,+\,\bar{b}.$\\ 
The signature of this presentation 
is a triple vertex with a leading $3\pom$ term.
The 3$\pom$ approximation is valid, when 
$\frac{m_p^2}{M_{sd}^2}\,<<\,1$ and 
$\frac{M_{sd}^2}{s}\,<<\,1$.\\ 
The leading energy/mass dependences are
\begin{equation} 
\frac{d\sigma^{3\pom}}{dt\,dM_{sd}^2} \propto 
s^{2\Delta_{\pom}} (\frac{1}{M_{sd}^2})^{1+\Delta_{\pom}}.
\end{equation}
Mueller's 3$\pom$ approximation for non GW  
diffraction is the lowest order of t-channel
multi $\pom$ interactions, 
which induce compatibility with t-channel unitarity.\\
Recall that unitarity screening of GW ("low mass") 
diffraction is carried out  
explicitly by eikonalization, while the  
screening of non GW ("high mass") diffraction is carried out 
by the survival probability (to be discussed).
\begin{figure*}
\includegraphics[width=90mm,height=30mm]{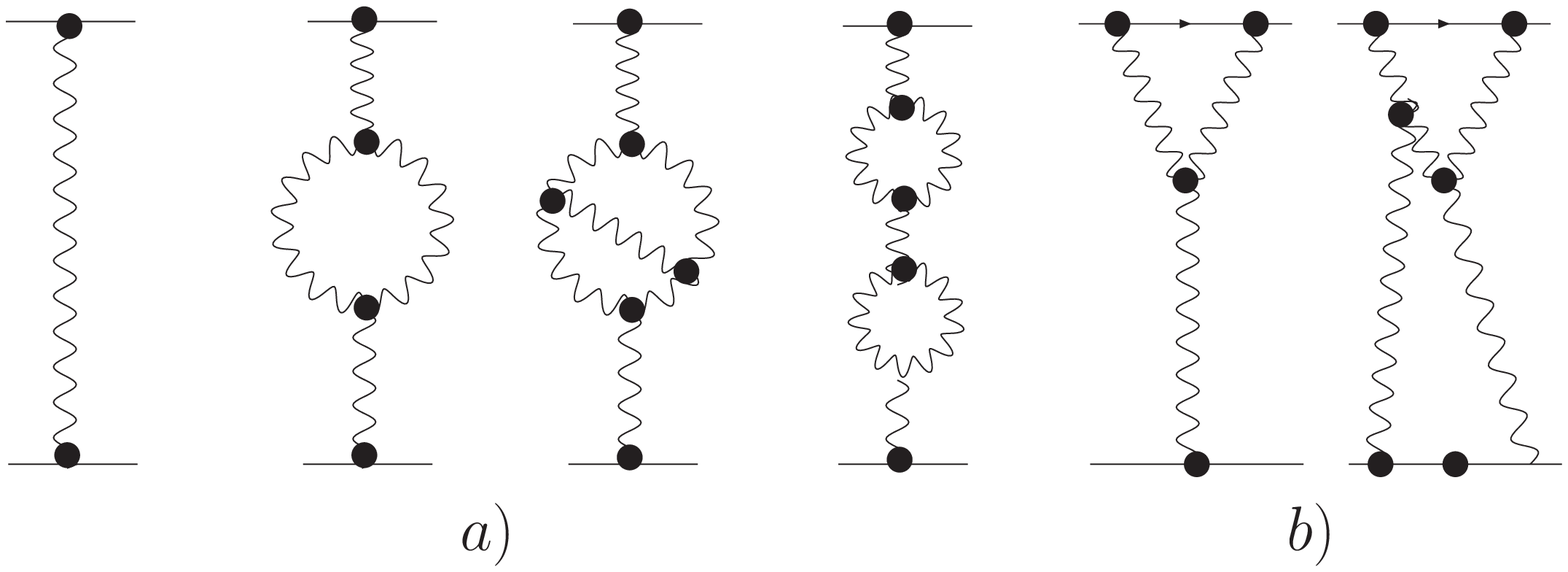}
\caption{The Pomeron Green function.}
\end{figure*}
Fig.3 shows the $\pom$ Green function.
Multi $\pom$ interactions are summed 
differently in the various $\pom$ models.\\
Note the analogy with QED renormalization:\\
a) Enhanced diagrams, 
present the renormalization of the propagator.\\ 
b) Semi enhanced diagrams,  
present the p$\pom$p vertex renormalization. 
\subsection{Survival Probability}
The experimental signature of a $\pom$ exchanged reaction
is a large rapidity gap (LRG), devoid of hadrons
in the $\eta-\phi$ lego plot,
$\eta\,=\,-ln(tan\frac{\theta}{2})$.\\
$S^2$, the LRG survival probability,
is a unitarity induced suppression factor
of non GW diffraction, soft or hard:
\begin{equation}   
S^2\,=\,{\sigma_{diff}^{screened}}/
{\sigma_{diff}^{nonscreened}}.
\end{equation}
It is the probability that the
LRG signature will not be filled by debris
(partons and/or hadrons) originating   
from either the s-channel re-scatterings of the spectator partons,
or by the t-channel multi $\pom$ interactions.\\
Denote the gap survival factor initiated by
s-channel eikonalization $S_{eik}^2$, and
the one initiated by
t-channel multi $\pom$ interactions, $S_{enh}^2.$\\
The eikonal re-scatterings of the incoming projectiles
are summed over (i,k).\\
$S^2$ is obtained from a convolution of
$S_{eik}^2$ and $S_{enh}^2.$
A simpler, approximation, is
\begin{equation}
S^2=S_{eik}^2 \cdot S_{enh}^2.
\end{equation}
\section{THE PARTONIC POMERON}
Current $\pom$ models differ in details, but have in common 
a relatively large adjusted input $\Delta_{\pom}$ and
a very small $\alpha_{\pom}^{\prime}$.
The exceedingly small fitted $\alpha_{\pom}^{\prime}$  
implies a partonic description 
of the $\pom$ which leads to a pQCD interpretation.\\
The microscopic sub structure of the $\pom$ is obtained from  
Gribov's partonic interpretation of Regge theory,
in which the slope of the 
$\pom$ trajectory is related to the mean
transverse momentum of the  
partonic dipoles constructing the Pomeron 
and, consequently, the running QCD coupling:
\begin{eqnarray}
\alpha_{\pom}^{\prime}\,&\propto&\,1/< p_t >^2,\nonumber \\
\alpha_S &\propto& \pi/ln \left(< p_t^2 >/\Lambda_{QCD}^2 \right) << 1.
\end{eqnarray}
We obtain a single $\pom$ with hardness depending on  
external conditions.\\
This is a non trivial relation as
the soft $\pom$ is a simple moving pole in J-plane,
while, the BFKL hard $\pom$ is a branch cut 
approximated, though, as a simple pole with  
$\Delta_{\pom}=0.2-0.3$, $\alpha_{\pom}^{\prime}=0$.
GLM and KMR models are rooted in Gribov's
partonic $\pom$ theory with a hard pQCD $\pom$ input.
It is softened by unitarity screening (GLM), or the 
decrease of its partons' transverse momentum (KMR).\\
Both models have a bound of validity, at 60(GLM) and 100(KMR) TeV, 
implied by their approximations. Consequently, as attractive as 
updated $\pom$ models are, we can not utilize them above 100 TeV.\\
To this end, the only available models are single channel, 
most of which have  
a logarithmic parametrization input. The main deficiency of such models 
is that while they provide a good reproduction of the 
available total and elastic data, 
their predictions at higher energies are questionable since 
diffractive channels and t-channel screening are not included 
\section{IS SATURATION ATTAINABLE? (PHENOMENOLOGY)} 
\subsection{Total and Inelastic Cross Sections:}
Table I compares $\sigma_{tot}$ and $\sigma_{inel}$ outputs of
GLM, KMR and BH in the energy range of 7-100 TeV.\\
Note that, GLM predictions at 100 TeV are above the model
validity bound.\\
As seen, the 3 models have compatible
$\frac{\sigma_{inel}}{\sigma_{tot}}$ outputs in the TeV-scale
which is significantly larger than 0.5.\\
The BH model can be applied at arbitrary high energies.
The prediction of BH at the Planck-scale (1.22$\cdot10^{16}TeV$) is,
$\sigma_{inel}/\sigma_{tot}=1131mb/2067mb=0.55,$
which is below $a_{el}$ saturation.
Recall, BH do not consider t-channel unitarity screening.
\subsection{$\Delta_{\pom}^{eff}$ Dependence on Energy}

\begin{table*} 
\begin{tabular}{|l|l|l|l|l|l} 
\hline 
&\,\,\,\,\,\,\,\,\,\,\,\,\,\,\,\,\,\,\,$7\,TeV$ &
\,\,\,\,\,\,\,\,\,\,\,\,\,\,\,\,\,\,\,$14\,TeV$ &
\,\,\,\,\,\,\,\,\,\,\,$57\,TeV$ &
\,\,\,\,\,\,\,\,\,\,\,\,\,\,\,\,\,\,$100\,TeV$ \\
&\, GLM\,\,\,\,KMR\,\,\,\,\,BH\,\, &
\,\,\, GLM\,\,\,\,KMR\,\,\,\,\,\,BH\,\, &
\, GLM\,\,\,\,\,\,\,\,\,BH\,\, &
\, GLM\,\, KMR\,\,\,\,\,\,BH\,\, \\
\hline
\,\,$\sigma_{tot}\,$ & \,\,\,\,\,98.6 \,\,\,\,\,97.4 \,\,\,\,95.4 &
\,\,\,\,109.0 \,\,\,107.5 \,\,107.3 & \,\,\,130.0 \,\,\,\,\,134.8 &
\,\,\,139.0 \,\,\,138.8\,\,\,\,\,147.1 \,\\
\hline
\,\,$\sigma_{inel}\,$ &
\,\,\,\,\,74.0 \,\,\, 73.6 \,\, 69.0 & \,\,\,\,\,\,\,\,81.1
\,\,\, 80.3
\,\,\,\,\,76.3 & \,\,\,\,\,\,95.2 \,\,\,\,\,\,\,92.9 & \,\,\,101.5
\,\,\,100.7\,\,\,\,\,\,100.0 \,\\
\hline
\,\,$\frac{\sigma_{inel}}{\sigma_{tot}}$ &
\,\,\,\,0.75 \,\,\,\,\,\,0.76 \,\,\,\,0.72 &
\,\,\,\,\,\,\,\,0.74 \,\,\,\,\,\,0.75\,\,\,\,\,\,\,0.71 &
\,\,\,\,\,\,0.73 \,\,\,\,\,\,0.70 &
\,\,\,\,\,\,0.73 \,\,\,\,\,\,0.73\,\,\,\,\,\,\,\,0.68 \\
\hline
\end{tabular}
\caption{Total and elastic cross sections}
~\\~\\~\\
\begin{tabular}{|l|l|l|l|l|l|l} 
\hline
\,\,\,\,\,\,\,\,\,\,\,\,
\,TeV\,&
\,\,1.8 $\rightarrow$\,\, 7.0 \,&
\,\,7.0 $\rightarrow$ 14.0 \,& 
\,7.0  $\rightarrow$ 57.0 \,&
\,57.0  $\rightarrow$ 100.0 \,&
\,14.0  $\rightarrow$ 100.0 \,\\
\hline
\,$\Delta_{eff}(GLM)$\,&\,\,\,\,\,\,\,\,\,0.081\,&\,\,\,\,\,\,\,\,0.072\,&
\,\,\,\,\,\,\,\,\,\,\,0.066\,&
\,\,\,\,\,\,\,\,\,\,\,0.060\,&\,\,\,\,\,\,\,\,\,\,\,\,\,0.062\,\,\\
\hline
\,$\Delta_{eff}(KMR)$\,&\,\,\,\,\,\,\,\,\,0.076\,&\,\,\,\,\,\,\,\,0.071\,&
\,\,\,\,\,\,\,\,\,\,\,\,&
\,\,\,\,\,\,\,\,\,\,\,\,&\,\,\,\,\,\,\,\,\,\,\,\,\,0.065\,\,\\
\hline
\,$\Delta_{eff}(BH)$\,&\,\,\,\,\,\,\,\,\,0.088\,&\,\,\,\,\,\,\,\,0.085\,&
\,\,\,\,\,\,\,\,\,\,\,0.082\,&
\,\,\,\,\,\,\,\,\,\,\,0.078\,&\,\,\,\,\,\,\,\,\,\,\,\,\,0.080\,\\
\hline
\end{tabular}
\caption{$\Delta_{\pom}^{eff}$ Dependence on Energy}
~\\~\\~\\
\begin{tabular}{|l|l|l|l|l|l}
\hline
&\,\,\,\,\,\,\,\,\,\,\,\,\,\,$7\,TeV$
&\,\,\,\,\,\,\,\,\,\,\,$14\,TeV$
&\,$57\,TeV$
&\,\,\,\,\,\,\,\,\,\,\,\,$100\,TeV$\\
&\,\, GLM\,\,\,\,\,KMR\,\,
&\,\, GLM\,\,\,\,KMR\,\,
&\,\,\,GLM\,\,
&\,\,\,\, GLM\,\,\,\,\,\,KMR\,\,\\
\hline
\,\,\,\,\,\,$\sigma_{tot}\,$ &
\,\,\,\,\,98.6 \,\,\,\,\,\,\,97.4 &
\,\,\,\,109.0 \,\,\,107.5 &
\,\,\,130.0 &
\,\,\,\,\,\,134.0 \,\,\,138.8 \\
\hline
\,\,\,\,\,\,$\sigma_{el}\,$ &
\,\,\,\,\,24.6 \,\,\,\,\, 23.8 &
\,\,\,\,\,\,27.9 \,\,\,\, 27.2 &  
\,\,\,\,\,\,34.8 &
\,\,\,\,\,\,\,\,37.5 \,\,\,\,\, 38.1 \\
\hline
\,\,\,\,\,\,$\sigma^{GW}_{sd}\,$ &
\,\,\,\,\,10.7 \,\,\,\,\,\,\,\,\,\,7.3 &
\,\,\,\,\,\,11.5 \,\,\,\,\,\,\,\,\,8.1 &
\,\,\,\,\,\,13.0 &
\,\,\,\,\,\,\,\,13.6 \,\,\,\,\,\,\,10.4 \\
\hline
\,\,\,\,\,\,$\sigma_{sd}\,$ &     
\,\,\,\,\,14.88 \,\,\,\,\,\,\,\,\,\,&
\,\,\,\,\,\,17.31 \,\,\,\,\,\,\,\,\,&  
\,\,\,\,\,\,21.68 &
\,\,\,\,\,\,\,\, \,\,\,\,\,\,\, \\
\hline
\,\,\,\,\,\,$\sigma^{GW}_{dd}\,$ &
\,\,\,\,\,\,\,6.21 \,\,\,\,\,\,\,\,\,0.9 &
\,\,\,\,\,\,\,\,6.79 \,\,\,\,\,\,\,\,1.1 &
\,\,\,\,\,\,\,\,7.95 &
\,\,\,\,\,\,\,\,\,\,\,8.39 \,\,\,\,\,\,\,1.6 \\
\hline
\,\,\,\,\,\,$\sigma_{dd}\,$ &
\,\,\,\,\,\,\,7.45 \,\,\,\,\,\,\,\,\,\,\,&
\,\,\,\,\,\,\,\,8.38 \,\,\,\,\,\,\,\, &
\,\,\,\,\,18.14 &
\,\,\,\,\,\,\,\,\,\,\, \,\,\,\,\,\,\,\\
\hline
\,\,$\frac{\sigma_{el}+\sigma_{diff}^{GW}}{\sigma_{tot}}\,$ &
\,\,\,\,\,\,0.42 \,\,\,\,\,\,\,0.33 &
\,\,\,\,\,\,\,0.42 \,\,\,\,\,0.34 &
\,\,\,\,\,\,0.43 &
\,\,\,\,\,\,\,\,0.43 \,\,\,\,\,\,\,\,\,0.36 \\
\hline
\end{tabular}
\caption{Diffractive cross sections}
\end{table*}
$\Delta_{\pom}^{eff}$ serves as a simple
measure of the rate of cross section growth estimated as 
$s^{\Delta_{\pom}^{eff}}.$
When compared with the adjusted input $\Delta_{\pom}$, we can assess 
the strength of the applied screening.\\
The screenings of
$\sigma_{tot}, \sigma_{el}, \sigma_{sd}$, $\sigma_{dd}$ 
and $M_{diff}^2$ are not 
identical. Hence, their $\Delta_{\pom}^{eff}$ values are different. 
The cleanest determination of $\Delta_{\pom}^{eff}$ is from the 
energy dependence of $\sigma_{tot}$. 
All other options require also 
a determination of $\alpha_{\pom}^{\prime}.$\\
Table II compares $\Delta_{\pom}^{eff}$ values obtained
by GLM, KMR and BH.
The continuous reduction of $\Delta_{\pom}^{eff}$ 
is a consequence of s and t screenings. 
\subsection{Diffractive Cross Sections}
GLM and KMR total, elastic and diffractive cross sections
are presented in Table III. 
KMR confine their predictions to the GW sector.\\
GLM GW $\sigma_{sd}$ and $\sigma_{dd}$
are larger than KMR. Their $\sigma_{tot}$ and $\sigma_{el}$
are compatible.\\
In both models, the GW components are compatible with the Pumplin bound.\\
The persistent growth of the diffractive cross sections
indicates that saturation will be attained (if at all)
well above the TeV-scale.\\
Analysis of diffraction, is hindered by different
choices of signatures and bounds!
\subsection{MacDowell-Martin Bound}
MacDowell-Martin Bound is
\begin{eqnarray} 
\frac{\sigma_{tot}}{B_{el}}
\leq 18 \pi\frac{\sigma_{el}}{\sigma_{tot}}. \nonumber
\end{eqnarray}
GLM and KMR ratios and bounds are:
\begin{eqnarray}
7~TeV:~~\frac{\sigma_{tot}}{B_{el}} &=& 12.5 < 14.1 (GLM), \nonumber \\
\frac{\sigma_{tot}}{B_{el}} &=& 12.3 < 13.8 (KMR). \nonumber \\
~ \nonumber \\
14~ TeV:~~
\frac{\sigma_{tot}}{B_{el}} &=& 13.0 < 14.5 (GLM),\nonumber \\
\frac{\sigma_{tot}}{B_{el}} &=& 12.8 < 14.3 (KMR).\nonumber \\
~ \nonumber \\
100~ TeV:~~
\frac{\sigma_{tot}}{B_{el}} &=& 13.8 < 15.3 (GLM), \nonumber \\
\frac{\sigma_{tot}}{B_{el}} &=& 13.8 < 15.5 (KMR). \nonumber
\end{eqnarray}
As seen, the ratios above are compatible with 
a non saturated $a_{el}(s,b)$ at the available energies.
\section{CONCLUSION}
The analysis presented re-enforced the critical roll played by 
s and t channel unitarity screenings in hadron-hadron high energy 
interactions. This presentation centered on p-p collisions at the 
TeV-scale with special attention invested on an 
assessment of unitarity saturation. Since the formalism of unitarity 
screenings is model dependent we have to be careful in the definitions 
of signatures indicating the onsetting of saturation.
\begin{itemize}
\item
A clear, model independent, saturation signature is 
\begin{equation}
\frac{\sigma_{el}}{\sigma_{tot}}\,=
\,\frac{\sigma_{inel}}{\sigma_{tot}}\,=\,\frac{1}{2}.
\end{equation}
Checking the experimental available cross section data,
leads to a definite conclusion that unitarity saturation in p-p 
scattering is NOT attained at the available energies. Checking the 
rate at which $\frac{\sigma_{inel}}{\sigma_{tot}}$ grows with energy, 
it is reasonable to conclude that saturation will not be attained at 
the TeV-scale and possibly (BH) up to the Planck-scale.
\item
Quite a few models confine their analysis exclusively to the p-p 
elastic channel. In my opinion, there is no way to bypass the 
coupling between the elastic and diffractive channels.
\item
Since diffraction cross sections vanish when unitarity saturation 
is attained,
we can consider that a change in the energy dependence of the diffractive 
cross section from a very moderate increase with energy to a decrease 
toward zero is an early signature that p-p scattering is approaching 
saturation. Since such a behavior has not been observed 
or predicted, I presume that saturation will not be attained at energies 
that can be experimentally investigated.
\end{itemize}
\end{document}